\documentclass[aps,twocolumn,preprintnumbers,superscriptaddress,prx]{revtex4-2}
\usepackage{graphicx}
\usepackage{amsmath}
\usepackage{amssymb}
\usepackage{color}
    
\begin{document}

\title{Single-mode lasing based on $\mathcal{PT}$-breaking of two-dimensional higher-order topological insulator}

\author{Bofeng Zhu}

\affiliation{Division of Physics and Applied Physics, School of Physical and Mathematical Sciences,\\
Nanyang Technological University, Singapore 637371, Singapore}

\affiliation{School of Electrical and Electronic Engineering, \\
Nanyang Technological University, Singapore 637371, Singapore}
  
\author{Qiang Wang}

\affiliation{Division of Physics and Applied Physics, School of Physical and Mathematical Sciences,\\
Nanyang Technological University, Singapore 637371, Singapore}

\author{Yongquan Zeng}

\affiliation{Electronic Information School, \\
Wuhan University, Wuhan 430072, China.}

\author{Qi Jie Wang}

\email{qjwang@ntu.edu.sg}

\affiliation{Division of Physics and Applied Physics, School of Physical and Mathematical Sciences,\\
Nanyang Technological University, Singapore 637371, Singapore}

\affiliation{School of Electrical and Electronic Engineering, \\
Nanyang Technological University, Singapore 637371, Singapore}
  
\affiliation{Centre for Disruptive Photonic Technologies, Nanyang Technological University, Singapore 637371, Singapore}

\author{Y.~D.~Chong}

\email{yidong@ntu.edu.sg}

\affiliation{Division of Physics and Applied Physics, School of Physical and Mathematical Sciences,\\
Nanyang Technological University, Singapore 637371, Singapore}

\affiliation{Centre for Disruptive Photonic Technologies, Nanyang Technological University, Singapore 637371, Singapore}

\begin{abstract}
  Topological lasers are a new class of lasers that seek to exploit the special properties of topological states of light.  A typical limiting factor in their performance is the existence of non-topological states with quality factors comparable to the desired topological states.  We show theoretically that by distributing uniform gain and loss on two sublattices of a two-dimensional higher-order topological insulator (HOTI) lattice, single-mode lasing based on topological corner states can be sustained over a wide range of pump strengths. This behavior stems from the parity/time-reversal breaking of the topological corner states, which supplies them with more effective gain than the edge and bulk states, rather than through localized pumping of the domain corners. These results point to opportunities for exploiting non-Hermitian phenomena and designing compact high performance topological lasers.
\end{abstract}

\maketitle

Photonic topological insulators (PTIs) are an emerging class of photonic devices based on adapting the topological band insulator concept, drawn from condensed matter physics, to light \cite{LuReview2014, OzawaReview2019, Haldane2008, Raghu2008, Wang2009}.  PTIs host novel photonic modes with unusual properties, such as robustness against disorder-induced localization, that may be useful for compact and low-loss waveguiding (e.g., in optical \cite{Hafezi2011, Hafezi2013}, quantum optical \cite{Barik2018}, and terahertz chips \cite{Singh2020}), signal processing (e.g., amplification and harmonic generation \cite{Peano2016, Kivshar2019}), lasing \cite{StJean2017, Longhi2018, Zhao2018, Parto2018, Ota2018, Harari2018, Bandres2018, Zeng2020, Gao2020, Yang2020, Zhang2020, Han2020, Kim2020, Zhong2021, Smirnova2020, Shao2020, Noh2020}, and other applications \cite{SmirnovaReview2020, OtaReview2020}.  Of these, PTI-based lasers have drawn particular attention.  They are conceptually intriguing since lasers are intrinsically non-Hermitian and nonlinear, whereas the theory of topological band insulators is formulated in the Hermitian and linear regime, raising the possibility of new phenomena arising from the interplay between band topology and laser physics \cite{Malzard2018, Longhi2018b, Kartashov2019, Secli2019, Secli2020, Amelio2020, Zapletal2020, Daniel2020}.  The first PTI lasers were based on one-dimensional (1D) Su-Schrieffer-Heeger (SSH) lattices, with SSH topological boundary states promoted to laser modes by the application of gain \cite{StJean2017, Zhao2018, Parto2018, Ota2018}.  Subsequently, PTI lasers were devised with two-dimensional (2D) lattices \cite{Harari2018, Bandres2018, Zeng2020, Dikopoltsev2021}, whose backscattering-resistant 1D topological edge states allow for robust single-mode lasing \cite{Harari2018, Bandres2018, PartoReview2021, Dikopoltsev2021}.  Lasers based on higher-order topological insulator (HOTI) lattices have also been studied \cite{Zhang2020, Han2020, Kim2020, Zhong2021}.  Unlike first-order topological insulators, HOTIs manifest their topological states on higher-order boundaries such as the ``zero-dimensional'' corners of 2D domains \cite{Benalcazar2017HOTI, Schindler2018HOTI}.  Photonic HOTIs in general \cite{Garcia2018, Mittal2018, Jiang2020, Ben2019, Chen2019, Hassan2019, Xie2020}, and HOTI lasers in particular \cite{Zhang2020, Han2020, Kim2020, Zhong2021}, are a promising avenue of research since they can be relatively easy to fabricate as photonic crystal slabs, and to integrate into existing photonic platforms.

PTI lasers of all types face the challenge of ensuring that the topological edge or boundary states lase preferentially.  There is usually no \textit{a priori} reason for topological states to have higher Q-factors than the bulk states, so simple pumping schemes tend to induce multimode lasing, with much of the emission caused by bulk modes. Previous 2D PTI laser designs have sought to address this problem by selectively pumping the spatial regions where the topological states reside \cite{Harari2018, Bandres2018, Zeng2020}; in HOTI lasers, for instance, corner regions can be selectively pumped to induce lasing of topological corner states \cite{Zhang2020, Han2020, Kim2020, Zhong2021}.  Another approach is to use frequency-selective gain \cite{Secli2020}, but this is harder to implement in semiconductor lasers.  A third approach, which has been employed in 1D SSH lasers \cite{Zhao2018, Parto2018}, is to apply gain to only part of each unit cell (e.g., one of the two sites in each SSH unit cell); even if this pumping scheme is applied uniformly throughout the lattice, the asymmetric gain distribution can supply more effective gain to the topological state than the bulk states \cite{Zhao2018, Parto2018, Schomerus2013}.  Thus far, this last type of pumping scheme has not been explored beyond 1D lattices.

In this paper, we show that in a HOTI laser containing gain and loss in different sublattices, the lasing of topological states can be protected by parity/time-reversal ($\mathcal{PT}$) breaking \cite{Schomerus2013, Longhi2017, feng2017, El-Ganainy2017}. The 2D HOTI lattice exhibits stable single-mode lasing over a substantial span of pump strengths.  The laser mode, composed of a superposition of topological corner states determined by mirror symmetry and $\mathcal{PT}$ symmetry breaking, experiences higher effective gain than all competing edge and bulk states, even without localized pumping of the domain corners.  By contrast, a standard localized pumping scheme (pumping the domain corners without exploiting $\mathcal{PT}$ breaking) induces multimode lasing, with the topological corner states not even having the lowest threshold.  The use of $\mathcal{PT}$ breaking to stabilize non-topological single-mode lasing has previously been demonstrated \cite{Feng2014, Hodaei2014}, and the present work shows that single-mode lasing based on topological HOTI corner states can be protected in a similar way, through the suppression of competing edge and bulk states.  This points to further opportunities for using the combination of non-Hermitian effects and band topology, a subject of considerable recent theoretical interest \cite{Bergholtz2021}, in the design of future topological laser devices.

\begin{figure}
  \centering
  \includegraphics[width=0.48\textwidth]{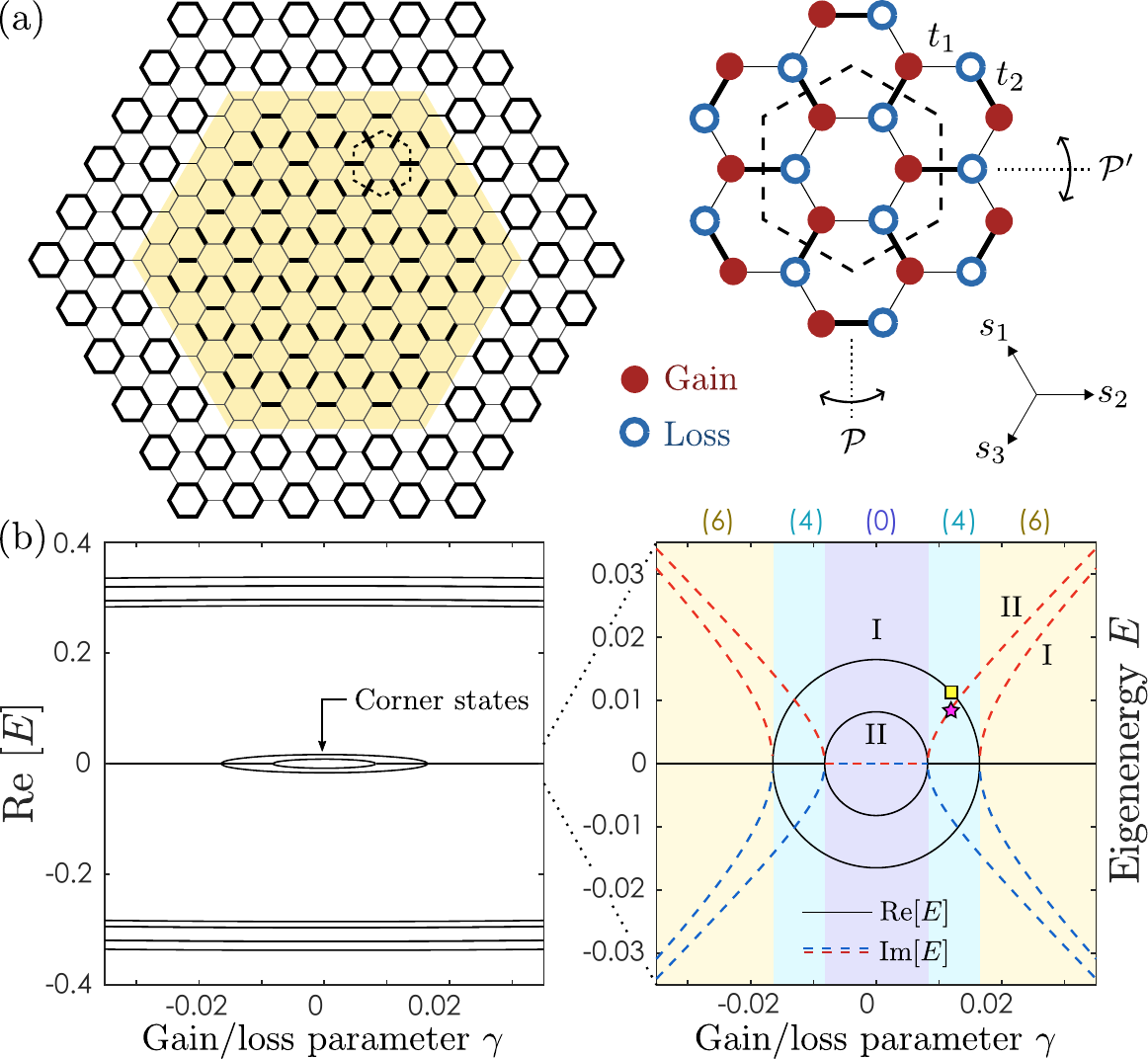}
  \caption{(a) Schematic of the lattice.  A hexagonal inner domain (yellow) is surrounded by an outer domain (white).  In the Hermitian and linear regime, the inner domain is a higher order topological insulator (HOTI) and the outer domain is trivial.  In the close-up view (right panel), dashes mark out one unit cell, thin (thick) lines indicate intracell (intercell) couplings $t_1 = 1-\delta$ ($t_2 = 1+\delta$), sites with gain (loss) are filled red (unfilled blue) circles, and the reflection symmetry operations $\mathcal{P}$ and $\mathcal{P}'$ are indicated.  (b) Linear spectrum, plotted against the gain/loss parameter $\gamma$ for $\delta = 0.6$.  Left panel: $\mathrm{Re}[E]$ versus $\gamma$.  At $\gamma = 0$, there are six mid-gap corner states, which undergo $\mathcal{PT}$ breaking as $|\gamma|$ is increased.  Right panel: zoomed-in plot of $\mathrm{Re}[E]$ (solid curves) and $\mathrm{Im}[E]$ (dashes) versus $\gamma$.  The branches labelled $\mathrm{I}$ and $\mathrm{II}$ are one- and two-fold degenerate respectively.  Square and star markers indicate the modes in Fig.~\ref{fig:spect_latt}.  The number of non-real eigenvalues in each $\gamma$ range is shown in parentheses.}
  \label{fig:schematic_latt}
\end{figure}

Our analysis is performed using a tight-binding model containing nearest-neighbor couplings, with each site representing an optical resonator (a similar approach has been successfully used to model real PTI lasers \cite{Zhao2018, Harari2018, Shao2020, Smirnova2020}). The model is both non-Hermitian and nonlinear: the pumped lattice sites are modelled with saturable gain dependent on the local site intensity, while the remaining lattice sites are subject to linear fixed loss (representing a combination of outcoupling loss and material absorption).  The above-threshold behavior of the laser is studied using both frequency-domain eigenmode analysis and time-domain simulations.

\begin{figure}
  \centering
  \includegraphics[width=0.48\textwidth]{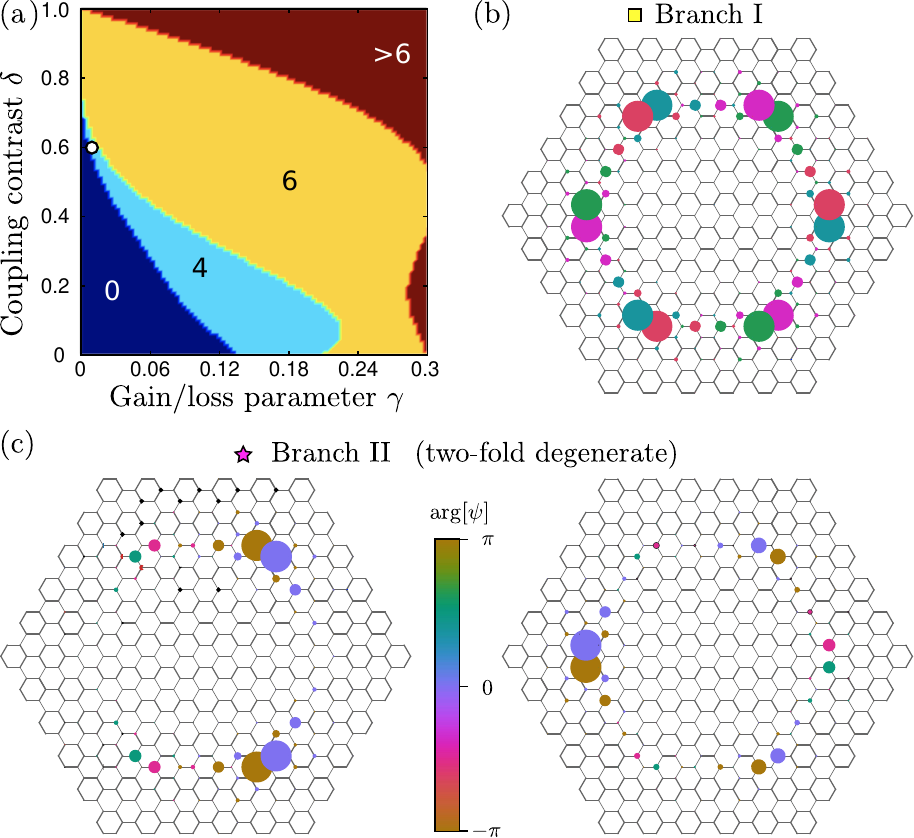}
  \caption{(a) Phase diagram showing the number of $\mathcal{PT}$-broken eigenstates versus $\gamma$ and $\delta$. The white dot indicates a representative point with parameters $\gamma = 0.012$ and $\delta = 0.6$, used in other subplots. (b)--(c) Intensity and phase distribution of selected eigenstates, corresponding to the markers in Fig.~\ref{fig:schematic_latt}(b).  The area of the circles is proportional to the intensity, and the color indicates the phase.  (b) Eigenstate on the non-degenerate branch I, which is $\mathcal{PT}$ symmetric (real $E$) and antisymmetric under $\mathcal{P}'$.  (c) Eigenstates on the degenerate branch II, which are $\mathcal{PT}$ broken with $\mathrm{Im}(E) > 0$ (their damped $\mathcal{PT}$ partners are not shown), and $\mathcal{P}'$-symmetric (left) and $\mathcal{P}'$-antisymmetric (right) respectively.}
  \label{fig:spect_latt}
\end{figure}

\begin{figure*}
  \centering
  \includegraphics[width=0.85\textwidth]{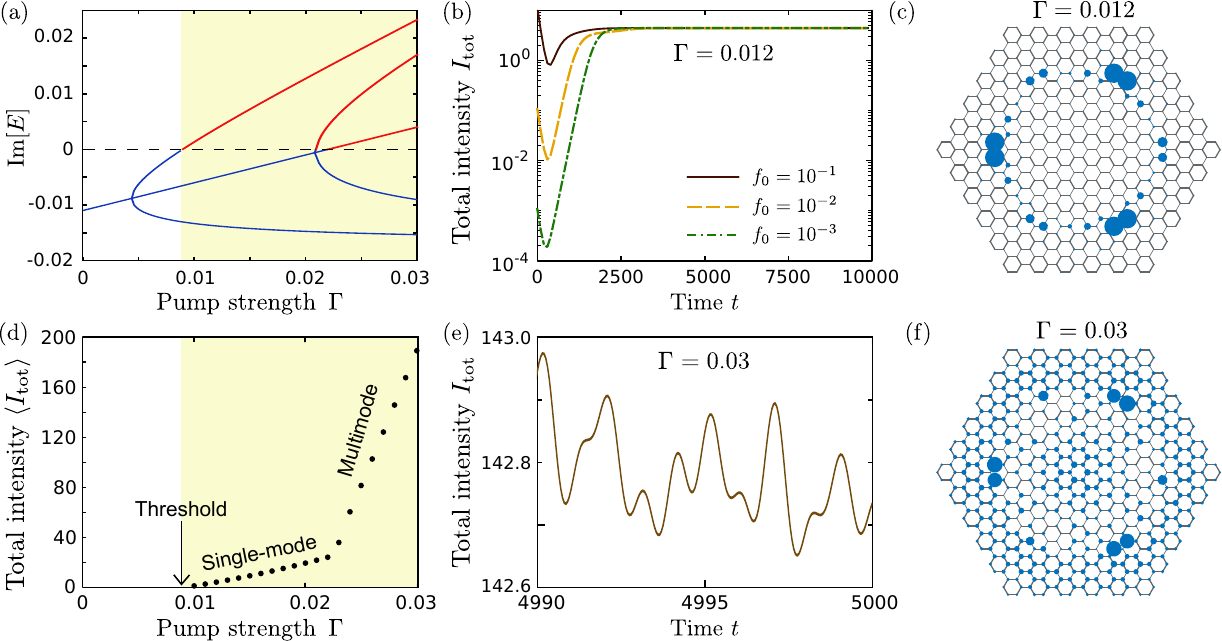}
  \caption{(a) Plot of $\mathrm{Im}[E]$ versus pump parameter $\Gamma$ for the corner states, in the linear regime (no gain saturation).  Lasing occurs upon crossing $\mathrm{Im}[E] = 0$ (horizontal dashes); the above-threshold regime is marked in yellow.  In this and subsequent subplots, we take $\gamma_0 = 0.005$, $\gamma_B = 0.017$, and $\delta = 0.6$.  (b) Time dependence of the total intensity $I_{\mathrm{tot}} = \sum_n |\psi_n|^2$, with $\Gamma = 0.012$ and gain saturation enabled. Each site is initialized to $(\alpha + i \beta) f_0$, where $\alpha $ and $ \beta $ are random numbers drawn from the standard normal distribution and the constant $f_0$ determines the magnitude of the initial excitation; results are shown for three independent initial conditions with different $f_0$, showing convergence to the same steady state $I_{\mathrm{tot}}$. (c)  Intensity distribution in the steady state.  A comparison with Fig.~\ref{fig:spect_latt}(c) shows that the lasing mode is a superposition of $\mathcal{P}'$-symmetric and $\mathcal{P}'$-antisymmetric modes with broken $\mathcal{PT}$.  (d) Time-averaged total intensity $\left\langle I_{\mathrm{tot}}\right\rangle$ versus $\Gamma$.  The kink at $\Gamma \approx 0.022$ corresponds to the onset of multimode lasing.  The $\left\langle I_{\mathrm{tot}}\right\rangle$ values are obtained by averaging over $t \in [9000, 10000]$. (e) Time dependence of $I_{\mathrm{tot}}$ in the multimode regime, $\Gamma = 0.03$. For this $t$ range, the transients have died away and the laser is in steady state. (f) Snapshot of intensity distribution in the multimode regime, taken at $t = 5000$ with the same parameters as in (e). In (c)--(f), the system is initialized with $f_0 = 0.01$, but the same results are obtained for other choices of $f_0$.}
  \label{fig:eigenvect_dyn}
\end{figure*}

The model, depicted in Fig.~\ref{fig:schematic_latt}(a), is a Wu-Hu lattice \cite{Wu2015} with gain/loss added to alternate sites.  There are six sites per unit cell, and the intracell (intercell) couplings are $t_1 = 1-\delta$ ($t_2 = 1 + \delta$), where $\delta$ is a real tuning parameter.  In a real system, such as a photonic crystal or waveguide array, $\delta$ may be set by the ratio of distances between inter- and intra-cell structural elements \cite{Wu2015, Garcia2018, Mittal2018}. In the inner (outer) domain, $\delta > 0$ ($\delta < 0$).  In the absence of gain/loss, the inner domain is a higher-order topological insulator (HOTI) in the nontrivial topological phase \cite{Ben2019,Xie2020}, surrounded by a topologically trivial cladding, so topological corner states will occur at each of the six corners of the domain.  Similar to previously-studied HOTI lasers \citep{Zhang2020, Han2020, Kim2020, Zhong2021}, the higher-order topology is tied to the ``filling anomaly'' of the $C_{6v}$-symmetric topological crystalline insulator \cite{Ben2019, Xie2020}.  Next, we subdivide the sites into two sublattices, A (gain) and B (loss), drawn as red and blue circles in the right panel of Fig.~\ref{fig:schematic_latt}(a).  In the position representation, the Hamiltonian is
\begin{multline}
  H= \left\{ \sum_{r \in A}^{} \sum_{i=1}^{3} t_{r,s_i}a_r^{\dagger}b_{r+s_i}
  + \mathrm{h.c.} \right\}  \\
  + \sum_{r \in A}^{} i \gamma_A(r)\; a_r^{\dagger}a_r
  - \sum_{r \in B}^{} i \gamma_B \, b_r^{\dagger}b_r,
  \label{finite_Ham}
\end{multline}
where $a_r$ ($b_r$) denotes the annihilation operator on the A (B) site at position $r$; $s_{1,2,3}$ are nearest neighbor displacement vectors; $t_{r,s_i} \in [t_1,t_2]$ are nearest neighbor hoppings assigned as stated above; $\gamma_{A}(r)$ is a gain term on each A site, to be discussed below; and $\gamma_{B}$ is a loss term for the B sites.  Each eigenvalue of $H$, denoted by $E$, represents a detuning relative to a principal frequency.

We first analyze the system in the absence of gain saturation and with balanced gain/loss $\gamma_A = \gamma_B = \gamma$.  The lattice is $\mathcal{PT}$ symmetric \cite{El-Ganainy2017}, where $\mathcal{P}$ is a reflection around the vertical axis as indicated in Fig.~\ref{fig:schematic_latt}(a), and $\mathcal{T}$ is complex conjugation.  Fig.~\ref{fig:schematic_latt}(b) plots the eigenvalue spectrum versus $\gamma$ for $\delta = 0.6$.  At $\gamma = 0$, there are six eigenstates in two sub-bands near $E = 0$, which are the six HOTI corner states; there is some hybridization-induced splitting due to the finite separation between corners.  As $|\gamma|$ is increased, the corner states successively undergo $\mathcal{PT}$ breaking \cite{Schomerus2013, Longhi2017, feng2017, El-Ganainy2017}: the eigenvalues coalesce at $\mathrm{Re}[E] = 0$ and become imaginary, as shown in the right panel of Fig.~\ref{fig:schematic_latt}(b). The two-fold degenerate branches labelled II break $\mathcal{PT}$ first; the non-degenerate outer branches labelled I break $\mathcal{PT}$ for larger $|\gamma|$.

Notably, the lattice's other eigenstates (i.e., bulk and edge states) have unbroken $\mathcal{PT}$ symmetry over the entire range of $\gamma$ shown here, and their eigenvalues are real.  In other words, \textit{the corner states undergo $\mathcal{PT}$ breaking first}.  In the lasing model to be discussed below, this is the key to producing laser modes from the topological corner states.  For much larger $|\gamma|$, the other eigenstates eventually break $\mathcal{PT}$ and acquire complex eigenvalues.  Fig.~\ref{fig:spect_latt}(a) shows the phase diagram for the number of non-real eigenvalues versus $\gamma$ and $\delta$. The $\mathcal{PT}$ breaking of the corner states depends not only on $\gamma$, but also on the effective coupling between the corners, which is determined by both $\delta$ and the system size.  In the infinite lattice limit where the corners do not couple, the $\mathcal{PT}$-breaking transition is thresholdless \cite{LiGe2014,SM}.

Aside from $\mathcal{PT}$ symmetry, the lattice is invariant under reflection across the horizontal axis, denoted by $\mathcal{P}'$.  The non-degenerate branch I is antisymmetric under $\mathcal{P}'$, and the two-fold degenerate branch II contains a $\mathcal{P}'$-symmetric and a $\mathcal{P}'$-antisymmetric mode.  The $\mathcal{P}'$ symmetry/antisymmetry persists through the $\mathcal{PT}$ breaking transition.  Fig.~\ref{fig:spect_latt}(b)--(c) shows three exemplary mode profiles at $\gamma = 0.012$ and $\delta = 0.6$, for which branch I is $\mathcal{PT}$ symmetric and branch II is $\mathcal{PT}$ broken.

We now develop the tight-binding model into a laser model by introducing gain saturation according to
\begin{equation}
  \gamma_{A} = \frac{\Gamma}{1+|\psi_n|^2} - \gamma_0.
  \label{gainsat}
\end{equation}
This describes a combination of saturable gain and background loss, where $\Gamma$  parameterizes the pump strength (taken to be the same on all A sites), $|\psi_n|^2$ is the intensity on site $n$, and $\gamma_0$ is a background loss.  Each B site is assigned a fixed loss $\gamma_B$.  The losses are due to some combination of outcoupling loss and material absorption, such that the laser's output power can be assumed to be proportional to the total intensity $I_{\mathrm{tot}} = \sum_n |\psi_n|^2$.

\begin{figure}
  \centering
  \includegraphics[width=0.48\textwidth]{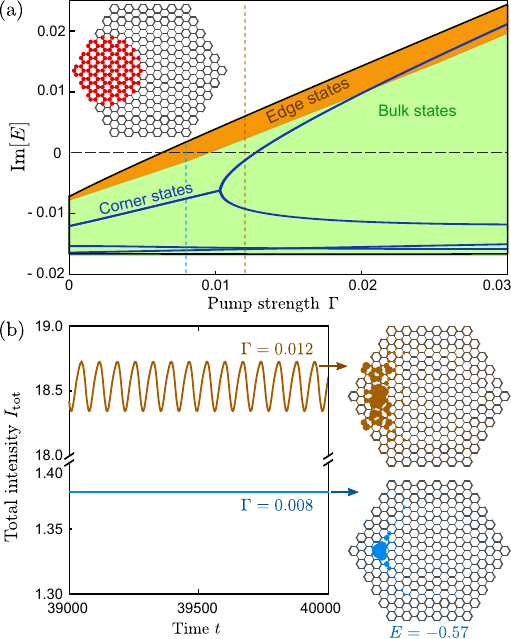}
  \caption{(a) Plot of $\mathrm{Im}[E]$ versus pump parameter $\Gamma$ for all eigenstates of a selectively pumped lattice, in the linear regime with no gain saturation.  The corner states are denoted by solid lines, and the areas marked in orange and green indicate where the edge and bulk states reside (the numerous individual edge and bulk states are not plotted for clarity).  Inset: schematic of the lattice with pumped sites marked by red circles.  The other sites all have loss $\gamma_A = \gamma_B = 0.017$.  In this and subsequent subplots, we take $\gamma_0 = 0.005$ and $\delta = 0.6$. (b) Time dependence of the total intensity $I_{\mathrm{tot}} = \sum_n |\psi_n|^2$ for $\Gamma = 0.008$ (blue curve) and $\Gamma = 0.012$ (brown curve), with gain saturation enabled. Each site is initialized to $(\alpha + i \beta) f_0$, where $\alpha, \beta$ are drawn from the standard normal distribution and $f_0 = 0.01$.  The right panels show snapshots of the intensity distributions taken at $t = 40000$.}
  \label{fig:partial_pump}
\end{figure}

For $|\psi_n|^2\rightarrow 0$, this is a linear lattice with $\mathcal{PT}$ symmetric gain/loss $\gamma = (\Gamma - \gamma_0 + \gamma_B)/2$ plus additional loss $\gamma' = (\gamma_B + \gamma_0 - \Gamma)/2$ on all sites.  Fig.~\ref{fig:eigenvect_dyn}(a) shows the plot of $\mathrm{Im}[E]$ versus $\Gamma$ for the corner states.  With increasing $\Gamma$, the corner states undergo bifurcations \cite{feng2017} that map to the previously-discussed $\mathcal{PT}$ breaking transitions. Note that these are unrelated to the bifurcations of nonlinear dynamical systems. The bulk and edge modes map to $\mathcal{PT}$ unbroken modes and have damping rate $\mathrm{Im}[E] = -\gamma'$.  The laser threshold occurs when one branch of corner states reaches $\mathrm{Im}[E] = 0$.

Above threshold, the gain saturation in Eq.~\eqref{gainsat} must be accounted for.  Fig.~\ref{fig:eigenvect_dyn}(b) shows time-domain simulations for the total intensity $I_{\mathrm{tot}}$ at pump strength $\Gamma = 0.012$.  The initial complex amplitude on each site $n$ is set to $\psi_n(t=0) = (\alpha_n + i \beta_n)f_0$, where $\alpha_n$ and $\beta_n$ are drawn independently from the standard normal distribution and $f_0$ is an overall scale factor for the initial state. The three curves in Fig.~\ref{fig:eigenvect_dyn}(b) corresponding to different $f_0$ all converge to the same $I_{\mathrm{tot}}$ at large times, indicating that the system operates as a stable single-mode laser.  Fig.~\ref{fig:eigenvect_dyn}(c) shows the steady state intensity profile.  Comparing this to Fig.~\ref{fig:spect_latt}, we see that the laser mode is a superposition of the $\mathcal{P}'$-symmetric and $\mathcal{P}'$-antisymmetric modes of branch II.  We verified numerically that the steady-state field distribution is indeed an equal-amplitude linear superposition of the two corner states in Fig.~\ref{fig:spect_latt}(c).

Fig.~\ref{fig:eigenvect_dyn}(d) shows the input-output curve (i.e., time-averaged total intensity versus pump strength $\Gamma$), obtained by time domain simulations.  The laser is single-mode up to about 2.4 times threshold ($\Gamma \approx 0.022$); the subsequent kink in the curve corresponding to the onset of multimode lasing. For these calculations, we take $f_0 = 0.01$ for the initial conditions and average $I_{\mathrm{tot}}$ over a time window $t \in [9000, 10000]$ (after transient oscillations have died away); we also verified that the resulting intensity curve is independent of $f_0$. Fig.~\ref{fig:eigenvect_dyn}(e) shows the complicated oscillations of $I_{\mathrm{tot}}$ in this regime.  Fig.~\ref{fig:eigenvect_dyn}(f) shows the intensity profile at a typical instant, confirming that the extra laser modes arise from bulk states; although the corner modes are visible, most of the output power is emitted from the more numerous bulk sites. The threshold and dynamic range of the single-mode lasing regime depends on $\delta$ and the system size, as discussed in the Supplemental Materials \cite{SM}.

To demonstrate the qualitative difference between the $\mathcal{PT}$-stabilized HOTI laser and a more standard design, Fig.~\ref{fig:partial_pump} shows the performance of a HOTI laser with a spatially selective pump.  The passive part of the HOTI lattice is the same as before, but all the sites near one domain corner are pumped, as illustrated in the inset of Fig.~\ref{fig:partial_pump}(a).  The plot of $\mathrm{Im}(E)$ versus pump strength $\Gamma$, in the absence of gain saturation, is shown in Fig.~\ref{fig:partial_pump}(a).  We see that the edge states and bulk states acquire effective gain comparable to and in some cases higher than the topological corner states.  Fig.~\ref{fig:partial_pump}(b) shows the time-dependent total intensity for two values of $\Gamma$; at $\Gamma = 0.008$, prior to the topological corner states reaching threshold, a mode evolving out of the edge state bands has already lased, as shown in the lower right panel of Fig.~\ref{fig:partial_pump}(b).  The mode frequency, determined via Fourier transformation, is $E = -0.57$, outside the band of topological corner states. At a slightly larger pump, $\Gamma = 0.012$, the lasing is multimode. Further time domain simulations reveal that single-mode lasing occurs over a relatively narrower range of pump strengths, as shown in the Supplemental Materials \cite{SM}. While this paper was in review, we noticed another paper concerning $\mathcal{PT}$-symmetry in a similar lattice \cite{Chen2021}, which focused on the $\mathcal{PT}$-breaking of the Bloch states rather than the topological corner states considered in this work.

We have thus demonstrated that by applying $\mathcal{PT}$ symmetric gain and loss to a HOTI lattice, topological corner states can be made to lase preferentially in a single mode over a wide dynamic range.  Competing edge and bulk states have significantly higher thresholds, in spite of spatial hole burning (local gain saturation), and without a frequency-selective gain medium.  The higher gain experienced by the topological laser mode arises from $\mathcal{PT}$ symmetry breaking, which has previously been shown to stabilize non-topological laser modes \cite{Schomerus2013, Longhi2017, feng2017, El-Ganainy2017}. Such a laser can be implemented with a 2D photonic crystal, as shown in the Supplemental Materials \cite{SM}. In future work, it would be interesting to study how other non-Hermitian symmetry breakings, and other inherently non-Hermitian phenomena, could be usefully incorporated into topological lasers.  


This work was supported by the Singapore MOE Academic Research Fund Tier 3 Grant MOE2016-T3-1-006, Tier 2 Grant MOE2018-T2-1-176, Tier 2 Grant MOE2016-T2-2-159, Tier 2 Grant MOE2019-T2-2-085, Tier 1 Grants RG187/18, and Singapore National Research Foundation (NRF) Competitive Research Program (CRP) (NRF-CRP18-2017-02).

\end{document}